# Multi-objective Quantum Annealing approach for solving flexible job shop scheduling in manufacturing

Philipp Schworm[a,*], Xiangquian Wu[a], Matthias Klar[a], Moritz Glatt[a], Jan C. Aurich[a]

[a] *Institute for Manufacturing Technology and Production Systems, RPTU Kaiserslautern, Germany*

*\*: Corresponding author, e-mail: [philipp.schworm@rptu.de](philipp.schworm@rptu.de), P.O. Box 3049, 67653 Kaiserslautern, Germany*

**Abstract**:

Flexible Job Shop Scheduling (FJSSP) is a challenging optimization problem with multiple conflicting objectives used to model and compute real-world process scheduling tasks. In order for a manufacturing system to remain competitive, it is necessary to compute such optimization problems quickly and efficiently. The limitations of conventional optimization methods frequently stem from a delicate balance between solution quality and computation time. Consequently, a pressing need for solution algorithms arises that can effectively transcend these limitations. This paper presents a novel Quantum Annealing-based solving algorithm (QASA) for computing FJSSP, leveraging the power of quantum annealing combined with classical techniques. The proposed approach aims to optimize a multi-criterial FJSSP considering makespan, total workload, and job priority simultaneously. QASA employs a Hamiltonian formulation with Lagrange parameters to integrate the problem's constraints and objectives. By assigning appropriate weights to the objectives, the method allows the prioritization of certain objectives over others. To handle the computational complexity of large FJSSP instances, the problem is decomposed into smaller subproblems, and a decision logic based on bottleneck factors is employed to select critical jobs for computation combined with variable pruning techniques. To evaluate the effectiveness of the proposed approach, experiments are conducted on benchmark problems, considering makespan, total workload, and priority objectives. Therefore, QASA combining tabu search, simulated annealing, and Quantum Annealing is used for efficient computation and is compared with a classical solving algorithm (CSA) combining tabu search and simulated annealing. The results demonstrate that QASA outperforms CSA in terms of solution quality, as measured by set coverage and hypervolume ratio metrics. Furthermore, computational efficiency analysis reveals that QASA achieves superior Pareto solutions compared to the classical approach, with a reasonable increase in computation time.

**Keywords**: process scheduling, job shop scheduling, multi-objective optimization, binary quadratic model (BQM), Quantum Annealing,

## Abbreviations

| | |
|---|---|
| BQM | Binary quadratic model |
| CSA | Classical solving algorithm |
| DJSSP | Dynamic job shop scheduling problem |
| FJSSP | Flexible job shop scheduling problem |
| HVR | Hypervolume ratio |
| JSS | Job shop scheduling |

| JSSP | Job shop scheduling problem |
| PPC | Production planning and control |
| QA | Quantum Annealing |
| QASA | Quantum Annealing-based solving algorithm |
| QPU | Quantum processing unit |

## 1. Introduction and Motivation

Production planning and control (PPC) is an essential function in manufacturing systems that involves managing the manufacturing and assembly processes through scheduling, capacity planning, and control of the production process. The primary objectives of PPC are to achieve on-time production and delivery, consistent capacity utilization, short lead times, low inventory, and high flexibility. These objectives are crucial to the economic viability of manufacturing systems, and efficient PPC is critical to achieving them [1]. However, the current global crises have resulted in significant market changes for manufacturing companies, necessitating effective approaches to manage these influences. These influences can manifest in changes in customer demands or sudden disturbances in supply chains. As the intensity and speed of these influences are unpredictable, approaches to manage them are required to ensure the economic viability of manufacturing system operations. Process scheduling is a crucial component of PPC that is responsible for scheduling and sequencing jobs in a manufacturing system [2]. The main task is to assign jobs to the various functional units such as machines or assembly stations according to the objectives of the manufacturing system [3]. However, the optimal assignment is complex due to numerous constraints, objectives, and possibilities, forming a challenging optimization problem known as the job shop scheduling problem (JSSP). This means that efficient and rapid optimization in response to the previously described unforeseen events is necessary, but challenging, especially for large job volumes. Due to the lengthy computation periods of traditional approaches such as exact methods or several heuristics, quick and effective optimization in reaction to unanticipated occurrences is inhibited, impeding the flexibility and responsiveness of the production system [4]. Thus, there is a need for computational methods that can provide efficient solutions to a multi-criterial JSSP in a short time frame.

In the field of algorithm research, meta heuristics are therefore increasingly used to solve such complex problems in a time-efficient manner [5]. Not only in the area of JSSP but also in wide areas of optimization. For example, Shaukat et. al maximize the life cycle and efficiency of a reactor core using a genetic algorithm coupled with monte carlo methods and achieve good results under a reasonable computational effort [6]. In addition, Lodewijks et. al minimize the total costs and energy consumption of the design of an airport baggage handling transport system using several particle swarm optimization algorithms. The results show that their algorithms can reduce $CO_2$ emission and costs within a short amount of time [7]. Moreover, such metaheuristics are used for many other optimization problems, such as the parameter estimation of solar cells [8]. Nevertheless, as problem sizes grow larger, metaheuristics are increasingly challenged by longer computation times and diminished solution quality. Consequently, the pursuit of more efficient algorithms in many research fields such as the JSSP remains a pressing and open area of research.

Recent research has highlighted the potential of quantum annealing (QA) in addressing this gap, as it can provide results within seconds for various intricate optimization problems [9,10]. QA is a computational approach that aims to discover the energy-minimum states of an optimization problem by utilizing quantum mechanical effects, such as superposition, entanglement, and tunneling, rather than merely simulating them, as done in simulated annealing [11]. Therefore, QA can enable the computation of solutions to combinatorial optimization problems efficiently, which may not be possible with other algorithms [12]. Previous studies have suggested that QA can offer rapid and effective solutions for flexible JSSPs (FJSSP), which are an extension of classical JSSPs where machines can be flexibly assigned to operations [13]. However, while the mono-criterial objective of this approach may be suitable for some applications, it can limit its applicability to real-world process scheduling in manufacturing systems. To address this limitation, an enhanced QA-based algorithm is developed for solving FJSSP that incorporates multicriterial problem formulations and an improved iterative approach to expedite computation time and achieve better solutions using bottleneck factors. The proposed approach is tested with a variety of objectives, including makespan, workload minimization, and priority, as combinations of two objectives and one multi-objective configuration with all three objectives. To evaluate the effectiveness of the proposed approach, a comparison is made with a simulated annealing algorithm for global search combined with a tabu search for local optimization. The comparison is conducted using modified MK problems from Brandimarte as benchmarks [14]. These MK problems include several FJSSPs with varying of jobs, processing times, and machines.

The following paper is structured as follows: Section 2 will highlight the state-of-the-art approaches of job shop scheduling (Subsection 2.1) and the basic principles of QA (Subsection 2.2). After presenting the research gap in Section 3, the methodology of the QA-based planning approach is described in Subsection 4.1 and Subsection 4.2, followed by the application scenarios (Subsection 4.3) that build the basis for the following investigations. Subsection 4.4 presents the mathematical formulation of the optimization model and all individual objectives followed by the multi-objective analysis and comparison with state-of-the-art algorithms in Subsection 4.5. The paper concludes with an outlook and summary.

## 2. State-of-the-art

### 2.1 Job shop scheduling approaches

Job shop scheduling (JSS) is an optimization problem from operations research and a widely used modeling technique for process scheduling tasks in manufacturing. There are many different types of problem formulations for JSSP. A distinction is made depending on the constraints and the object of consideration. The classical JSSP aims to allocate a set of given jobs, consisting of successive operations, to the machines of the manufacturing system under consideration of processing times in order to fulfill certain objectives [3]. Thereby, objectives can be grouped into time-based, job-number-based, cost-based, revenue-based, and energy- and environment-based criteria. The most widely used objective is the time-based objective makespan, i.e., the completion time of all jobs, to minimize the lead times of jobs in the manufacturing system. Moreover, several criteria can be considered individually or in combination as a multi-objective approach. In addition, constraints are considered to map the manufacturing system's conditions to the problem formulation [15]. Common constraints are a procedure constraint for the consideration of specified process sequences, an overlapping constraint to avoid multiple uses of a machine, and a processing constraint, which ensures that started operations are not interrupted and start multiple times [16]. Moreover, assumptions are made. Among others, these include the availability of machines all the time without consideration of failures and maintenance, jobs without priorities and due dates, neglection of transport and set-up times, and predetermined assignments of operations to exactly one machine [15]. However, since these assumptions and limitations lead to less accurate models of reality, different types of JSSPs are used to better represent real-world conditions by neglecting some assumptions. Flexible job shop scheduling (FJSSP) assumes that some machines can perform the same tasks and that operations can be variably assigned. This is associated with machine-dependent processing times and is used to ensure a more practical optimization [16]. Moreover, in dynamic job shop scheduling (DJSSP), time-based events such as machine failures and probabilistic job arrivals are considered [17]. Depending on the type of JSSP as well as the objectives, the problem complexity and the requirements for the planning algorithms differ. However, one characteristic that all types of JSSP have in common is the NP-hardness of the problem formulation [18]. In other words, finding an optimal solution for this problem takes a non-polynomial amount of time in relation to the problem size. As a result, it can be very time-consuming to obtain optimal solutions for medium-sized problems and almost impossible for large problem instances. The results of Brucker et al. show these effects by using a branch and bound algorithm for computing various problem instances of JSSP [19]. Consequently, approximation methods are used to compute solutions efficiently for larger optimization problems.

Approximation methods are categorized into constructive methods, artificial intelligence methods, local search algorithms, and meta-heuristics [4]. In order to solve various JSSP the optimization algorithms are used solely or in combination to exploit the specific advantages of the optimization methods. The different methods are thereby adapted according to the considered problem sizes and modeling techniques. Chakraborty et al. utilize a simulated annealing approach to calculate classical JSSP but acknowledge its limitations. To effectively tackle larger problem sizes, they suggest combining their approach with local search algorithms [20]. Furthermore, Zhang et al. employ a genetic algorithm to solve FJSSP but also acknowledge the aforementioned challenges [21]. Li et al. propose a solution for computing larger FJSSPs by combining a genetic algorithm with tabu search. Their approach demonstrates both scalability and high-quality results for well-known benchmark problems. However, the authors caution that multi-criteria considerations must be taken into account to satisfy the demands of a manufacturing environment [22]. However, adopting a multi-criteria approach results in a larger solution space, thereby increasing the problem size and the computational complexity required to obtain good solutions. Moreover, alternative evaluation methods are necessary to compare results. For instance, Caldeira et al. propose the discrete jaya algorithm for multi-objective FJSSP, which minimizes the objectives makespan, total workload, and critical workload [23]. To compare the results, performance metrics such as set coverage and hypervolume ratio are used. The approach generates solutions with diversity and convergence to the reference Pareto front. However, attaining benchmark solutions of comparable quality is associated with increasing computation times and a time-based termination criterion. Thus, the solution space is restricted and limits the algorithm in finding optimal solutions. In real applications, this would directly influence the performance indicators of a manufacturing system and thus the profits achieved. Another approach proposed by Gao et al. demonstrates that by considering makespan and the mean of earliness and tardiness in a multi-objective FJSSP, the MK benchmark problems of Brandimarte can be solved with fast computation times [24]. The results indicate that solutions can be found quickly and thus has a positive influence on the responsiveness of the manufacturing system. However, it is important to acknowledge that the suggested approach can only identify a restricted number of Pareto points in specific problem instances, thereby restricting the discovery of optimal solutions. As a result, tackling real-world applications with multiple objectives and complex problem instances can be challenging, particularly when working with weighted objectives. Therefore, there is a pressing need for computation methods that can deliver high-quality solutions while also achieving fast computation times. This is where QA can potentially offer a solution.

## 2.2 Quantum Annealing

QA is a quantum computing technique that leverages quantum mechanics to speed up solving combinatorial optimization problems [25]. QA can be used to find the minimum energy state of an optimization problem, which necessitates formulating the problem as an energy minimization problem. To represent an optimization problem as an energy minimization problem which can be computed by QA, for example, the notation as Hamilton function can be used. The Hamilton formulation maps an energy value to each state of an optimization problem and is characterized by binary variables $x_i, x_j$ and corresponding scalar weights $Q_{ii}, Q_{ij}$ (Eq. 1).

$$H = \sum_i Q_{ii} x_i + \sum_{i<j} Q_{ij} x_i x_j \tag{1}$$

Solving the Hamilton function requires mapping it onto the quantum processing unit which is a common requirement for any quantum algorithm that uses qubits. This mapping process is known as embedding and can be a challenge for quantum annealers due to the need for additional qubits and longer embedding times.

As the first company to develop and commercialize the quantum annealing system, D-Wave[1] has achieved exponential growth in qubit number from D-Wave One in 2011 with 128 qubits to the current D-Wave Advantage with 5640 qubits [26]. Moreover, it has developed various embedding techniques to optimize the minor embedding process. To save the complexities, auto embedding generates automatic mapping corresponding to different problems that can be used for simple problems and also cause long embedding times during changeable problems throughout the computation. Fixed embedding is a mapping strategy where the qubits and connections on the quantum chip are pre-mapped to the problem being solved, and the mapping remains fixed throughout the computation. Therefore, fixed embedding offers faster embedding times and may be more suitable for complex, large-scale problems. However, it may require more qubits and may not be as flexible as auto embedding. External embedding is another approach for mapping optimization problems onto the quantum annealer. In this strategy, an external solver or algorithm is employed to find the minor embedding of a problem, allowing for greater customization and flexibility in the embedding process. By utilizing specialized algorithms tailored for specific problem types or applying decomposition methods, external embedding can potentially result in more efficient embeddings and improved performance [27].

Despite the limitations in addressing extensive problems directly on the quantum processing unit (QPU) due to qubit constraints, a variety of hybrid solvers that integrate classical algorithms with quantum annealing can be used to address large-scale challenges. Furthermore, the open-source libraries for QA facilitate the specification of personalized hybrid workflows through the selection of diverse classical algorithms and embedding technologies, thus providing increased adaptability and customization. Owing to its ability to resolve intricate optimization issues, QA has found applications across various industry sectors such as finance [28–30], energy [31,32], transportation [33–35] and manufacturing [9,10]. In the context of JSSP, numerous investigations of QA have been conducted. Venturelli et al.'s proposed approach has demonstrated the feasibility of QA solving JSSP [36]. However, it must be noted that the problems under consideration are simplified small problem sizes where many boundary conditions of the real scenario have been neglected that are a long way from practical application. Despite, the potential of quantum-based methods for computation has been highlighted by Denkena et al., who employed a digital annealer to compute larger instances of the FJSSP [37]. The outcomes reveal not only the capability to handle larger problems but also offer advantages in terms of computation time and achieved solution quality. Nevertheless, since a digital annealer solely simulates the quantum mechanical effects of a quantum annealer, it is plausible to assume that utilizing a real quantum annealer could lead to better results.

---

[1] Naming of specific companies is done solely for the sake of completeness and does not necessarily imply an endorsement of the named companies nor that the products are necessarily the best for the purpose.

## 3. Research gap

Besides the previously mentioned application fields, recent developments enable the utilization of QA for a multi-criterial JSSP. To showcase the capabilities of quantum annealing in solving large-scale problems, previous studies have applied various hybrid solvers to address single-objective FJSSP of differing sizes, confirming QA's computational advantages over conventional methods [13,38]. However, most real-world planning problems require the consideration of multiple objectives. In order to make QA usable for such real-world applications, multi-criteria objectives have to be investigated. For multi-objective optimization problems, quantum-based algorithms have also been developed and shown to outperform classical algorithms [39]. However, the full potential of quantum algorithms for multi-objective optimization cannot be fully exploited if there are not implemented on the corresponding quantum hardware. Therefore, it is important to investigate whether these assumptions can be confirmed. For this reason, QA-based algorithms need to be developed that allow realistic objective combinations to be transferred to the FJSSP and computed, thus paving the way for QA to be used in real application scenarios.

## 4. Multi-objective flexible job shop scheduling using Quantum Annealing

### 4.1 Framework

In general, a model formulation is needed to optimize the allocation of jobs to the machines of a manufacturing system to fulfill certain objectives. Therefore, building a FJSSP in the proposed framework is the prerequisite. Figure 1 depicts a generalized approach for computing multi-objective FJSSP using QA. The initial step involves converting the machines $M$ and jobs $J$ of a manufacturing system into binary variables during pre-processing. These variables serve as the foundation for formulating mathematical constraints and objectives. Next, binary quadratic model (BQM) formulations for each constraint and objective are established by linking the variables with corresponding weights through analytical methods considering properties such as processing times. The Hamiltonian function is then formed by combining individual objective functions with scalar weighting factors (Lagrange parameters). The Lagrange parameters' values are determined based on the priorities of each objective, with constraints usually weighted higher than objective functions. Finally, the Hamiltonian function is embedded onto the quantum hardware and solved using QA-based algorithms. The resulting solution can be evaluated using performance metrics and visualization techniques.

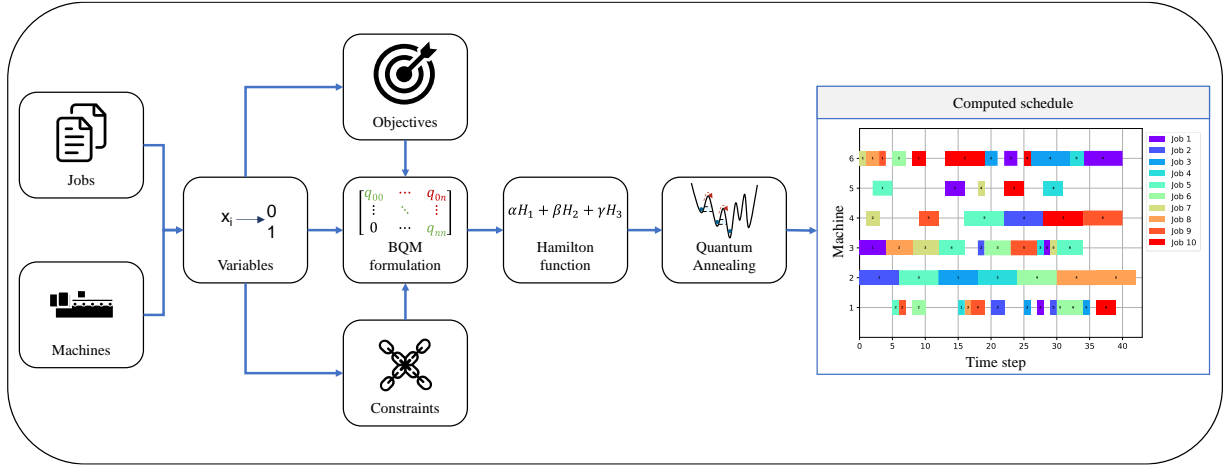

Figure 1: General framework for QA-based job shop scheduling

## 4.2 Mathematical formulation

Foundation of the proposed approach is a FJSSP determined through a number of given jobs $J = \{j_1, \ldots, j_A\}$ and a number of given machines $M = \{m_1, \ldots, m_B\}$. Each job $i \in J$ is presented through a predefined sequence of operations $O_i = \{o_{i1}, \cdots, o_{iC}\}$ and has a priority $pr_i \in PR$ which determines its urgency. An operation $o_{ij}$ can at least be processed on one machine $m_k \in M_{ij} \subseteq M$ within a corresponding processing time $p_{ijk} \in P = \{p_{111}, \ldots, p_{ACB}\}$. For the model formulation a discrete timeline $T = \{0, \ldots, T_{max}\}$ is defined in which all jobs must be scheduled. To determine the starting time $t$ of each operation $o_{ij} \in O_i$ binary variables are used. A variable $x_{ijkt} \in X$ is equal to 1 if operation $o_{ij}$ starts at time $t$ and is processed on machine $m_k$. Otherwise, the variable is equal to 0 (Eq. 2) [13].

$$x_{ijkt} = \begin{cases} 1: o_{ij} \text{ starts at time } t \text{ on machine } m_k \\ 0: otherwise \end{cases} \quad (2)$$

In the FJSSP, three constraints are considered. The *processing constraint $c_1$* causes each operation to be assigned to exactly one machine and starting at only one specific time step (Eq. 3). With the *procedure constraint $c_2$* the compliance of the predefined sequence of operations is ensured (Eq. 4). An *overlapping constraint $c_3$* ensures no multiple occupation of a machine by two different operations at the same time (Eq. 5-7) [13].

$$c_1 = H_4 = \sum_{i \in J} \sum_{o_{ij} \in O_i} \left(1 - \sum_{t \in T} \sum_{m_k \in M} x_{ijkt}\right)^2 \quad (3)$$

$$c_2 = H_5 = \sum_{\substack{i \in J \\ }} \sum_{\substack{o_{ij}, o_{ij'} \in O_i \\ j < j'}} \sum_{\substack{t'-t < p_{ijk} \\ (m_k, m_{k'}) \in M_{ij} \times M_{ij}}} x_{ijkt} \cdot x_{ij'k't'} \quad (4)$$

$$c_3 = H_6 = \sum_{m_k \in M_{ij} \cap M_{i'j'}} \sum_{\substack{(o_{ij}, o_{i'j'}) \in O_i \times O_{i'} \\ (i,i',t,t') \in G \cup H}} x_{ijkt} \cdot x_{i'j'kt'} \quad (5)$$

where

$$G = \{(i, i', t, t'): i, i' \in J, i \neq i', t, t' \in T, 0 \leq t - t' < p_{i'j'k}\} \quad (6)$$
$$H = \{(i, i', t, t'): i, i' \in J, i \neq i', t, t' \in T, 0 \leq t' - t < p_{ijk}\} \quad (7)$$

The constraints are represented as penalty functions using binary variables within a Hamiltonian. When a constraint is violated, the corresponding penalty function increases the energy value of the Hamiltonian. In principle, the consideration of further constraints is conceivable, e.g., including constraints related to machine unavailability, which would restrict start times to specific machine-allocated time slots. However, since these additional constraints are not taken into account in the current approach, it is established that the mentioned constraints are sufficient for the generation of valid results. The proposed approach encompasses multiple objectives, each of which holds significant relevance in the context of industrial planning. The primary objective is to minimize the makespan, which inherently contributes to the reduction of lead times in real-world operations. (Eq. 8-9). For implementation, any operation with a completion time exceeding the minimum predecessor time is subjected to a penalty. This minimum predecessor time is calculated by summing the minimum processing times of the preceding operations [13].

$$f_1 = H_1 = \sum_{\substack{i \in J \\ o_{ij} \in O_i \\ m_k \in M_{o_{ij}} \\ t \in T}} x_{ijkt} \cdot \left(t + p_{ijk} - P_{o_{ij}}\right) \tag{8}$$

where

$$P_{o_{ij}} = \sum_{j'<j} \min_{m_k \in M_{ij'}} p_{ij'k'} \tag{9}$$

The second objective function strives to minimize the total workload across all machines, leading to greater flexibility in job scheduling within real industrial companies. In this case, solutions that lead to longer durations of machine occupation are penalized. In order to do so, each binary variable is combined with the corresponding workload in a penalty function, so that low workloads are preferred. The workload is determined through the processing times on the machines (Eq. 10).

$$f_2 = H_2 = \sum_{i \in J} \sum_{o_{ij} \in O_i} \sum_{m_k \in M_{ij}} \sum_{t \in T} x_{ijkt} \cdot (p_{ijk} - \min_{m_{k'} \in M_{ij}} p_{ijk'}) \tag{10}$$

The third objective leads to an allocation of jobs according to a priority. Within an industrial context, such prioritization has the potential to expedite the processing of urgent orders, potentially resulting in reduced penalty costs and enhanced customer loyalty. Jobs with high priority have to be processed before jobs with lower priority if they compete for the occupancy of a machine at a certain time. To formulate this as a penalty function, an approach similar to the makespan objective is used where completion times that exceed the minimum predecessor time are penalized. In addition, a priority value is multiplied with the weight according to priority of the job (Eq. 11-12).

$$f_3 = H_3 = \sum_{\substack{i \in J \\ o_{ij} \in O_i \\ m_k \in M_{ij}}} x_{ijkt} \cdot \left(t + p_{ijk} - P_{o_{ij}}\right) \cdot pr_i \tag{11}$$

where

$$P_{o_{ij}} = \sum_{j'<j} \min_{m_{k'} \in M_{ij'}} p_{ij'k'} \qquad (12)$$

The constraints and objectives are combined in a Hamiltonian with corresponding non-negative Lagrange parameters $\alpha, \beta, \gamma, \delta, \varepsilon, \zeta$. In this way a BQM is created (Eq. 13).

$$H = \alpha \cdot H_1 + \beta \cdot H_2 + \gamma \cdot H_3 + \delta \cdot H_4 + \varepsilon \cdot H_5 + \zeta H_6 \qquad (13)$$

The Lagrange parameters determine the influence of the corresponding Hamilton term. This means that constraints usually have to be weighted higher than objectives in order not to obtain infeasible solutions. Furthermore, not all objectives have to be considered at the same time. Individual objectives can be neglected by setting the corresponding Lagrange parameter to zero. Thus, it is possible to consider all three objectives simultaneously or only two objectives together as well as a single objective function.

### 4.3 Iterative approach

The complexity of most multi-objective planning problems requires to limit their size to fit the limited capacity of quantum annealers. One approach is to break down the problem into smaller subproblems that can be computed iteratively. This involves analyzing only a specific number of operations $O^s$ of a subset of jobs $J_s$ at a time instead of the entire set. The outcomes of these subproblems are then combined to produce an overall result. However, to execute this strategy, a decision logic is needed to determine which job combinations should be allocated first. Bottleneck factors form the basis of this logic, enabling the selection of the most critical jobs for computation. In the context of the multi-objective FJSSP, the bottleneck factor for each job $i$ in each loop, denoted as $\delta_i^{(loop)}$ (Eq. 14), is conceived as the square root of the sum of the squared values of individual bottleneck factors corresponding to each considered objective, with each factor appropriately weighted.

$$\delta_i^{(loop)} = \sqrt{\alpha_l (\delta_{i,f_1}^{(loop)})^2 + \beta_l (\delta_{i,f_2}^{(loop)})^2 + \gamma_l (\delta_{i,f_3}^{(loop)})^2} \qquad (14)$$

In equation (2), $\delta_{i,f_1}^{(loop)}$, $\delta_{i,f_2}^{(loop)}$, and $\delta_{i,f_3}^{(loop)}$ represent the bottleneck factors for each objective $f_1, f_2, f_3$, and $\alpha_l, \beta_l$, and $\gamma_l$ denote their respective weights.

In this paper, three objectives are considered, which are makespan, total workload and priority. Therefore, the corresponding bottleneck factors are illustrated as follows.

- The bottleneck factor of job $i$ for the makespan objective, $\delta_{i,f_1}$ (Eq. 15), is calculated by subtracting the sum of minimum processing times for all remaining operations $O'_{i'}$ across all remaining jobs $J'$ from the minimum processing time of the remaining tasks $O'_i$ for job $i$. This difference is then normalized by the range of total minimum processing times for all remaining operations across all remaining jobs.

$$\delta_{i,f_1} = \frac{\sum_{o_{ij} \in O'_i} \min_{m_k \in M_{ij}} p_{ijk} - \min_{i' \in J'} \sum_{o_{i'j'} \in O'_{i'}} \min_{m_{k'} \in M_{i'j'}} p_{i'j'k'}}{\max_{i' \in J'} \sum_{o_{i'j'} \in O'_{i'}} \min_{m_{k'} \in M_{i'j'}} p_{i'j'k'} - \min_{i' \in J'} \sum_{o_{i'j'} \in O'_{i'}} \min_{m_{k'} \in M_{i'j'}} p_{i'j'k'}} \qquad (15)$$

- For the bottleneck factor of workload minimization, $\delta_{i,f_2}$ (Eq. 17) is calculated by defining machine bottlenecks $\delta_{i,m_k}$ (Eq. 16) for each operation $o_i$ in job $i$ in combination with possible machines $m_k$. The machine bottlenecks indicate how many operations of all remaining jobs can potentially be processed on the respective machine. This value is set in relation to the total number of remaining operations and summed with a term that reflects the number of potential machines of an operation. In order to calculate the bottleneck factors $\delta_{i,f_2}$, the machine bottlenecks $\delta_{i,m_k}$ are summarized over all operations of a job and normalized by dividing the maximum summarized machine bottlenecks over all jobs.

$$\delta_{i,m_k} = \frac{1}{\sum_{m_k \in M_{ij}} 1} + \frac{\sum_{\substack{i \in J' \\ o_{ij} \in O'_i \\ m_k \in M_{ij}}} \sum_{\substack{i' \in J' \\ o_{i'j'} \in O'_{i'} \\ m'_k \in M_{i'j'} \\ m_k = m'_k}} 1}{\sum_{\substack{i \in J' \\ o_{ij} \in O'_i}} 1} \tag{16}$$

$$\delta_{i,f_2} = \frac{\sum_{o_{ij} \in O'_i} \delta_{i,m_k}}{\max_{i' \in J'} \sum_{o_{i'j'} \in O'_{i'}} \delta_{i,m_k}} \tag{17}$$

- The bottleneck factor according to priority $\delta_{i,f_3}$ (Eq. 18) is defined by the priority values of the corresponding jobs. Each bottleneck factor is normalized by the maximum priority value of all jobs.

$$\delta_{i,f_3} = \frac{pr_i}{\max_{i' \in J} (pr_{i'})} \tag{18}$$

The adjustment of $\alpha_l$, $\beta_l$, and $\gamma_l$ determines the weighting of the objectives, resulting in a varied selection of jobs within a computation loop. This adjustment directly influences the selection of jobs based on the specific objective at hand. In cases where only two or a single objective is considered, the corresponding weightings can be set to zero to exclude that particular objective from consideration.

In addition to a job selection logic, several variable pruning methods are needed. In each loop, unfeasible solutions can be eliminated in order to minimize the optimization problem. Therefore, the previous defined predecessor time $P_{o_{ij}}$ is utilized which despites the minimum duration of the preceding operations of the current operation $o_{ij}$. Furthermore, a successor time $S_{o_{ij}}$ is defined (Eq. 19). The successor time is calculated as the sum of the minimum processing durations of all subsequent operations, including the processing duration of the current operation.

$$S_{o_{ij}} = \sum_{j' \geq j} \min_{m'_k \in M_{ij'}} p_{ij'k'} \tag{19}$$

By utilizing the predecessor and successor times, it becomes possible to establish a time range within which an operation has to be processed (Eq. 20). This time range is determined based on an estimated time interval $T_{max}$ specified in each computation iteration. Consequently, variables within the BQM that fall outside this interval can be eliminated, effectively reducing the complexity of the optimization problem.

$$T = \left[P_{o_{ij}}, T_{max} - S_{o_{ij}}\right] \tag{20}$$

Moreover, additional variables can be eliminated when considering solutions of previous computation loops. Thus, the predecessor time can be increased accordingly to the point in time at which the previous operation can start at the earliest without violating a constraint. Furthermore, in order to minimize the problem size for all solvers, it is advisable to choose a small value for the parameter $T_{max}$. This decision necessitates making an estimation based on factors such as the number of operations in the given jobs, available machines, and their corresponding processing times considered in one loop. By summing up the processing times of operations within each job in a loop, the maximum processing time across all jobs can establish a lower bound for $T_{max}$ in the loop. Similarly, the sum of processing times across all jobs can provide an upper bound for $T_{max}$. To ensure suitable values for varying problem sizes, $T_{max}$ can be incremented based on the number of operations and their processing durations, while being decreased with respect to the number of machines according to Eq. 21 with $T_{est}$ as a predetermined minimum time interval. The factor represented by $a_1$ signifies the proportion of jobs within the respective loop compared to the total sum of all jobs (Eq. 22), while $a_2$ denotes the ratio of operations associated with potential machine assignment conflicts relative to the total number of machines (Eq. 23).

$$T_{max} = \sum_{\substack{i \in J_s \\ o_{ij} \in O_i^S}} \max_{m_k \in M_{ij}} p_{ijk} + T_{est} \cdot a_1 \cdot a_2 \tag{21}$$

$$a_1 = \frac{\sum_{\substack{i \in J_s \\ o_{ij} \in O_i^S}} 1}{\sum_{\substack{i \in J \\ o_{ij} \in O_i}} 1} \tag{22}$$

$$a_2 = \frac{\sum_{\substack{i \in J_s \\ o_{ij} \in O_i^S \\ m_k \in M_{ij}}} \sum_{\substack{i' \in J_s \\ o_{i'j'} \in O_{i'}^S \\ m_{k'} \in M_{i'j'} \\ m_k = m_{k'}}} 1}{\sum_{\substack{i \in J_s \\ o_{ij} \in O_i^S \\ m_k \in M_{ij}}} 1} \tag{23}$$

The implementation of the previous described bottleneck factors and variable pruning method leads to a workflow shown in Figure 2.

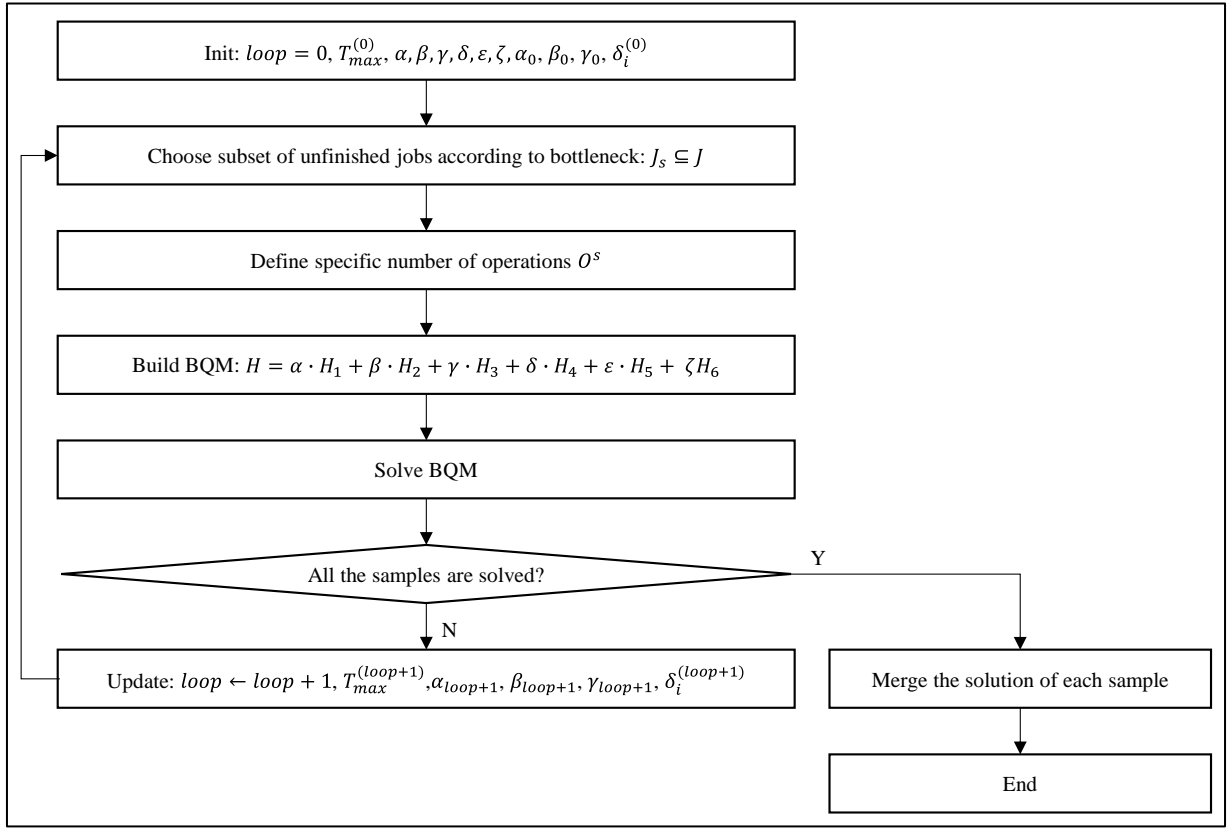

Figure 2: Iterative computation framework

## 4.4 Use case

The workflow described above presents a method for computing the FJSSP using a QA-based solving algorithm (QASA). However, due to the larger sizes of widely used FJSSP benchmarks, relying solely on a QPU seems not suitable due to the limited qubits number of the current quantum annealers. Using only a QPU would lead to numerous iterations and lower solution quality, as the tiny subproblems would be computed without considering the entire problem. To address this challenge, hybrid solvers that combine classical resources and QA are employed. These hybrid solvers allow for the consideration of larger problem instances in a single iteration, resulting in better solution quality while still benefiting from the fast computation provided by QA. The QASA in this study is built using a combination of tabu search, simulated annealing, and a QA-based implementation with an external embedding technique [40]. To evaluate the approach, 10 benchmark problems from Brandimarte et al. are utilized, with the addition of randomly assigned priorities to the jobs [14]. This enables the evaluation of the priority objective using the Brandimarte benchmark, not only considering the total workload and makespan. Furthermore, employing this benchmark simplifies future comparisons with the selected approach, as it has already been widely utilized as a reference point by numerous other methodologies. To ensure meaningful results, all problem combinations are also computed using a classical solving algorithm (CSA) for comparison purposes. CSA employs a simulated annealing and tabu search algorithm and follows the same workflow, including variable pruning techniques, bottleneck factor-based job selection, and identical constraints and objective formulation. This ensures that the comparison of algorithmic performance remains independent of the model formulation or the workflow, focusing solely on the algorithms themselves. To obtain comprehensive results, full factorial trials are conducted by varying different Lagrange parameters and job subset sizes. This procedure helps generate different Pareto points, which are then used as a basis for comparison between the two approaches. The investigations are divided into successive stages. The initial stage involves solving a multi-criterial problem by considering the minimization of makespan and total workload. In the second stage makespan minimization and priority is considered. In the third stage all objectives are used in combination. Table 1 displays a partial set of computation inputs utilized in the study to build the BQM. In the table, each entry represents an interval, with the lower bound denoting the first parameter and the upper bound indicating the second parameter for the corresponding value. As the valid input parameter ranges vary across different instances of the problem, this approach assists in determining appropriate parameter ranges based on the characteristics of each problem instance. Previous research findings suggest that constraints should be prioritized over objectives, and therefore, the value bounds are selected to be higher accordingly [13]. In the use case, when performing iterative computation, $\alpha_l$, $\beta_l$, and $\gamma_l$ are set to 1 if the corresponding objective is taken into consideration, and to 0 otherwise. This way, all objectives are considered equally when selecting job subsets, but their priority is controlled by the weightings in the Hamiltonians. The computations for both the QASA and CSA are performed on an Intel XEON_SP_6126 with 20GB RAM. In addition, the D-Wave Advantage is used as the quantum resource in QASA.

Table 1: Computation inputs of the BQM

| Objective | $T_{est}$ | $\alpha$ | $\beta$ | $\gamma$ | $\delta$ | $\varepsilon$ | $\zeta$ | $J_s$ | $O^s$ |
|---|---|---|---|---|---|---|---|---|---|
| $f_1 + f_2$ | [10,50] | [0.1,100.0] | [0.1,100.0] | 0 | [100,1500] | [100,1500] | [100,1500] | [2,10] | [2,10] |
| $f_1 + f_3$ | [10,50] | [0.1,100.0] | 0 | [0.1,100.0] | [100,1500] | [100,1500] | [100,1500] | [2,10] | [2,10] |
| $f_1 + f_2 + f_3$ | [10,50] | [0.1,100.0] | [0.1,100.0] | [0.1,100.0] | [100,1500] | [100,1500] | [100,1500] | [2,10] | [2,10] |

### 4.5 Multi-objective analysis with comparison

Each solution can be characterized by the value achieved in relation to the respective objective, which allows the use of different metrics for evaluation. In this way, makespan and total workload are evaluated by comparison of different makespan values $e_{f_1}$ or total workload $e_{f_2}$ values. In order to be able to apply the same comparison criterion to $f_3$, a corresponding value $e_{f_3}$ is defined (Eq. 24). Since earlier completion of a lower-priority job does not necessarily influence the early completion of a higher-priority job, a comparison of jobs based solely on priority is not appropriate. For this reason, makespan is also included in the evaluation. To determine $e_{f_3}$, the start time of the last operation of a job $t_{o_{iC}}$ added with the processing time of chosen machine $m_k$ is set in relation to the makespan and multiplied by a normalized priority value. The values are summed up over all jobs. Thus, the goal of minimization is directly reflected by this factor, as jobs with high priorities and short completion times are weighted low, while high priorities and long completion times are weighted high.

$$e_{f_3} = \sum_{i \in J} \frac{t_{o_{iC}} + p_{iCk}}{e_{f_1}} \cdot \frac{pr_i}{\max_{i' \in J}(pr_{i'})} \tag{24}$$

In order to evaluate multiple objectives at the same time, two evaluation metrics for solution quality are used. The first one is a set coverage metric (C-metric). This metric compares two sets of Pareto points (A, B) and makes statements about a superiority of a one set of Pareto points over the other (Eq. 25).

$$C(A,B) = \frac{|\{b \epsilon B \; \exists \; a \epsilon A : a \; dominates \; b\}|}{|B|} \tag{25}$$

If C(A, B) represents the percentage of solutions in B that are dominated by at least one solution in A, and C(A, B) yields a higher value than C(B, A), then it can concluded that the solutions of algorithm A outperform the results of algorithm B [41].

In addition, hypervolume ratio (HVR) is used for comparison of Pareto solutions (Eq. 26). The HVR is defined as the ratio of the hypervolume dominated by a set of solutions to the hypervolume of the entire objective space with $v_{pf}$ as the hypercube formed between a solution in the obtained Pareto front. The hypervolume is calculated by considering the reference point, which represents the worst achievable values for each objective. The HVR provides a measure of how well a set of solutions covers the objective space or how close it is to the optimal Pareto front. A higher HVR indicates a better spread and coverage of the objective space by the solutions, implying a higher quality set of solutions. Conversely, a lower HVR suggests that the solutions are more concentrated or do not cover the objective space well [42].

$$\text{HVR} = \frac{volume(\bigcup_{pf=1}^{PF} v_{pf})}{volume(\bigcup_{pf'=1}^{PF'} v_{pf'})} \tag{26}$$

To evaluate the proposed approach, the computation times, along with the mentioned metrics, are documented. In case of QASA, an additional annealing time is recorded for distinctions between use of local resources and quantum hardware. This also creates comparability in terms of performance by including a time component in the evaluation. Therefore, the mean value of all computation times connected to a Pareto solution are determined for each problem instance. The pre- and post-processing times are neglected as they do not differ due to the same problem formulations as BQM workflows. To ensure fair comparisons, for each instance and objective, QASA and CSA are executed independently 100 times. The results of the computation are summarized in Table 2 and Figure 3-6. The computing times listed in Table 2 represent mean values, each derived from all run times and accompanied by its standard deviation.

Table 2: Computation results

| Problem instance | Objective | QASA | | | | CSA | | |
|---|---|---|---|---|---|---|---|---|
| | | $C(A,B)$ | HVR | Annealing time (s) | Computing time (s) | $C(B,A)$ | HVR | Computing time (s) |
| MK01 | $f_1 + f_2$ | 0.571 | 1.102 | 0.320±0.155 | 22.98±2.23 | 0.142 | 1.020 | 17.81±0.97 |
| | $f_1 + f_3$ | 0.670 | 2.605 | 0.396±0.125 | 19.96±4.47 | 0.100 | 1.263 | 17.00±0.00 |
| | $f_1 + f_2 + f_3$ | 0.368 | 0.966 | 0.468±0.131 | 22.78±2.18 | 0.167 | 0.840 | 17.85±2.62 |
| MK02 | $f_1 + f_2$ | 0.667 | 1.160 | 0.719±0.277 | 30.72±3.9 | 0.000 | 0.806 | 33.46±2.76 |
| | $f_1 + f_3$ | 0.636 | 4.139 | 0.918±0.553 | 37.08±4.47 | 0.222 | 2.359 | 25.30±3.43 |
| | $f_1 + f_2 + f_3$ | 0.474 | 3.015 | 0.458±0.153 | 32.30±3.85 | 0.231 | 2.320 | 23.39±4.127 |
| MK03 | $f_1 + f_2$ | 0.570 | 2.170 | 1.974±0.628 | 82.23±8.24 | 0.540 | 2.240 | 89.77±17.78 |
| | $f_1 + f_3$ | 0.412 | 3.707 | 1.083±0.328 | 81.34±10.83 | 0.235 | 3.014 | 110.50±18.5 |
| | $f_1 + f_2 + f_3$ | 0.222 | 1.557 | 1.062±0.393 | 53.54±6.51 | 0.111 | 1.030 | 70.07±13.86 |
| MK04 | $f_1 + f_2$ | 0.500 | 1.594 | 0.678±0.219 | 29.2±3.22 | 0.330 | 1.483 | 38.68±4.99 |
|  | $f_1 + f_3$ | 0.777 | 1.568 | 0.770±0.181 | 31.73±3.06 | 0.375 | 3.161 | 42.53±5.04 |
|  | $f_1 + f_2 + f_3$ | 0.654 | 1.760 | 0.642±0.127 | 32.18±3.78 | 0.043 | 0.490 | 33.39±3.85 |
| MK05 | $f_1 + f_2$ | 0.545 | 1.933 | 1.065±0.243 | 50.06±2.84 | 0.364 | 1.533 | 54.00±6.44 |
| MK05 | $f_1 + f_3$ | 0.461 | 3.077 | 0.935±0.332 | 37.33±5.78 | 0.220 | 2.413 | 65.78±3.39 |
|  | $f_1 + f_2 + f_3$ | 0.541 | 1.188 | 0.816±0.162 | 54.30±5.55 | 0.200 | 0.589 | 57.31±5.42 |
| MK06 | $f_1 + f_2$ | 0.500 | 2.983 | 1.033±0.220 | 70.47±4.55 | 0.3125 | 1.417 | 83.99±6.23 |
| MK06 | $f_1 + f_3$ | 0.538 | 4.207 | 1.008±0.460 | 78.49±9.46 | 0.312 | 2.878 | 109.21±12.94 |
|  | $f_1 + f_2 + f_3$ | 0.364 | 2.553 | 0.986±0.456 | 72.84±11.02 | 0.176 | 1.169 | 84.34±12.89 |
| MK07 | $f_1 + f_2$ | 0.714 | 2.115 | 0.886±0.095 | 53.08±5.56 | 0.091 | 0.832 | 73.93±5.45 |
| MK07 | $f_1 + f_3$ | 0.454 | 1.506 | 0.863±0.166 | 87.71±8.38 | 0.333 | 1.397 | 124.86±18.73 |
|  | $f_1 + f_2 + f_3$ | 0.606 | 0.511 | 0.960±0.307 | 68.00±8.77 | 0.109 | 0.531 | 68.41±9.43 |
| MK08 | $f_1 + f_2$ | 0.583 | 2.716 | 1.396±0.183 | 117.14±13.07 | 0.384 | 1.692 | 122.41±12.45 |
| MK08 | $f_1 + f_3$ | 0.917 | 2.425 | 1.342±0.575 | 151.33±10.04 | 0.059 | 2.377 | 164.05±17.29 |
|  | $f_1 + f_2 + f_3$ | 0.368 | 1.218 | 1.520±0.301 | 107.27±14.15 | 0.125 | 0.564 | 113.12±10.48 |
| MK09 | $f_1 + f_2$ | 0.692 | 2.124 | 1.747±0.226 | 113.93±9.16 | 0.307 | 1.188 | 127.57±17.72 |
| MK09 | $f_1 + f_3$ | 0.533 | 2.864 | 1.342±0.293 | 151.33±22.34 | 0.235 | 2.039 | 164.05±17.48 |
|  | $f_1 + f_2 + f_3$ | 0.162 | 2.060 | 1.902±0.887 | 125.27±18.29 | 0.136 | 1.370 | 130.24±14.72 |
| MK10 | $f_1 + f_2$ | 0.583 | 3.790 | 1.612±0.212 | 128.38±12.3 | 0.430 | 1.320 | 147.58±19.63 |
| MK10 | $f_1 + f_3$ | 0.571 | 4.879 | 1.691±0.730 | 98.89±8.07 | 0.077 | 3.206 | 177.81±22.91 |
|  | $f_1 + f_2 + f_3$ | 0.303 | 1.250 | 1.793±0.540 | 107.00±6.14 | 0.133 | 2.290 | 133.70±15.54 |

By analyzing the results, it can be concluded that QASA consistently outperforms CSA in terms of the C-metric across all problem instances and objective configurations. In 100 % of the cases, QASA achieves superior Pareto solutions compared to CSA. The overview of the C-metric comparison can be found in Figure 3. Here the C-metrics of QASA and CSA over all problem instances are visualized. In addition, the solutions of MK09 are exemplarily shown for $f_1 + f_2$ and $f_1 + f_3$ in Figure 4. The solutions are normalized with respect to the worst achieved Pareto solutions as reference points. Thus, in the case of $f_1 + f_2$ the dominance of QASA over CSA is directly apparent. Moreover, QASA outperforms CSA in terms of HRV in 84 % of the cases. Figure 5 provides a visualization of this comparison. It is worth noting that differences in the metrics arise due to QASA's ability to achieve superior Pareto solutions, resulting in the removal of other non-dominant solutions from the Pareto front. Consequently, the hyper volume, which measures the distribution of Pareto solutions, may exhibit lower values. This effect can be observed in MK07 and MK10 with the objective $f_1 + f_2 + f_3$. In terms of computational efficiency, Figure 6 presents the computing time for the approaches. For small problem sizes such as MK01 and MK02, CSA demonstrates better computing time compared to QASA. However, the difference in computation times is relatively small (e.g., 8,9 % higher for MK02 with objective $f_1 + f_2$). Conversely, as problem sizes increase, QASA surpasses CSA in terms of computation time. This holds true for all larger problem sizes and, overall, 84% of all problem instances. Furthermore, it is worth noting that the annealing times for both approaches fall within the range of 0.3 to 2.0 seconds across all problem sizes, indicating relatively low values compared to the total computation time (less than 2%). Differences in computation time can also be attributed to the type of access to quantum hardware. While CSA solely utilizes local resources, QASA accesses quantum hardware via the cloud. Consequently, the total computing time for QASA can be divided into local computing time, transmission time to and from the quantum annealers, embedding time, and access time (actual annealing time). The variations in total computing time can be explained by this allocation, with the transmission time assumed to be nearly constant and its proportion of the total computing time decreasing with larger problem sizes due to higher processing unit demand.

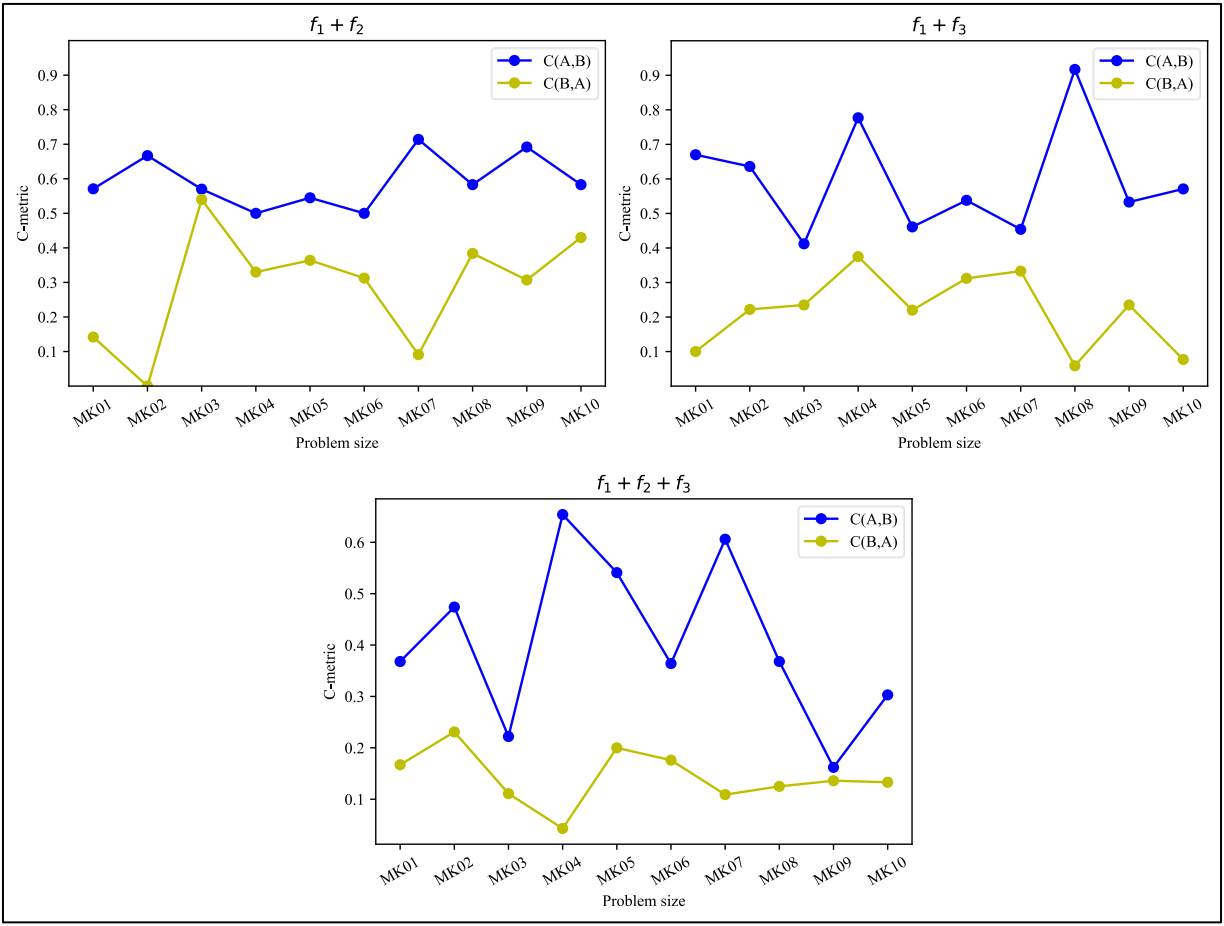

Figure 3: Comparison of C-metric

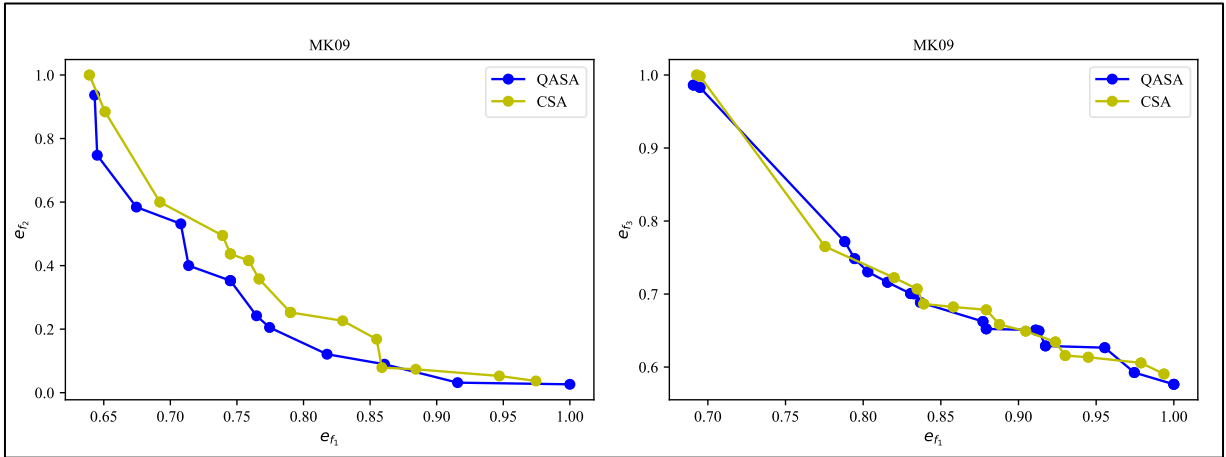

Figure 4: Comparison of Pareto points of MK09

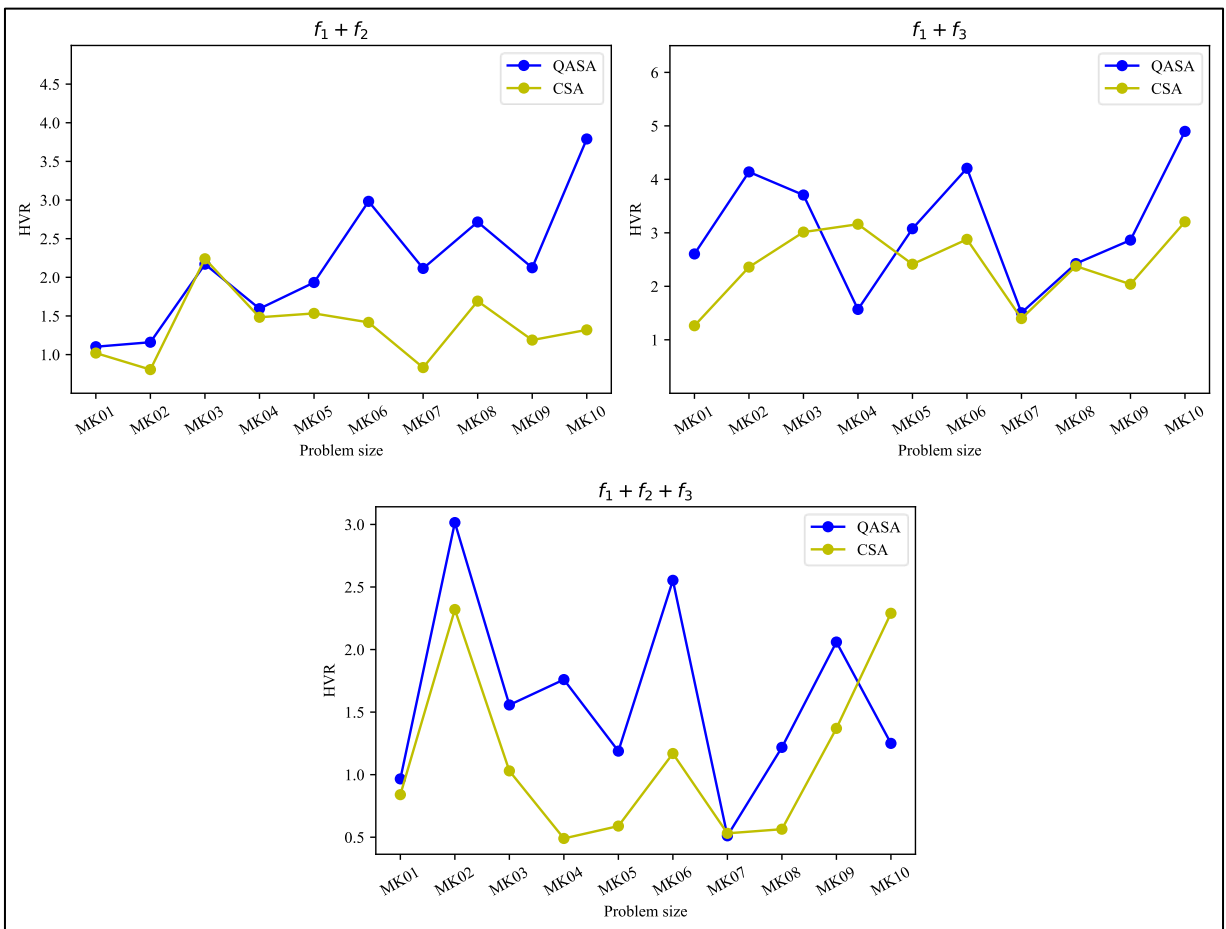

Figure 5: Comparison of HVR metric

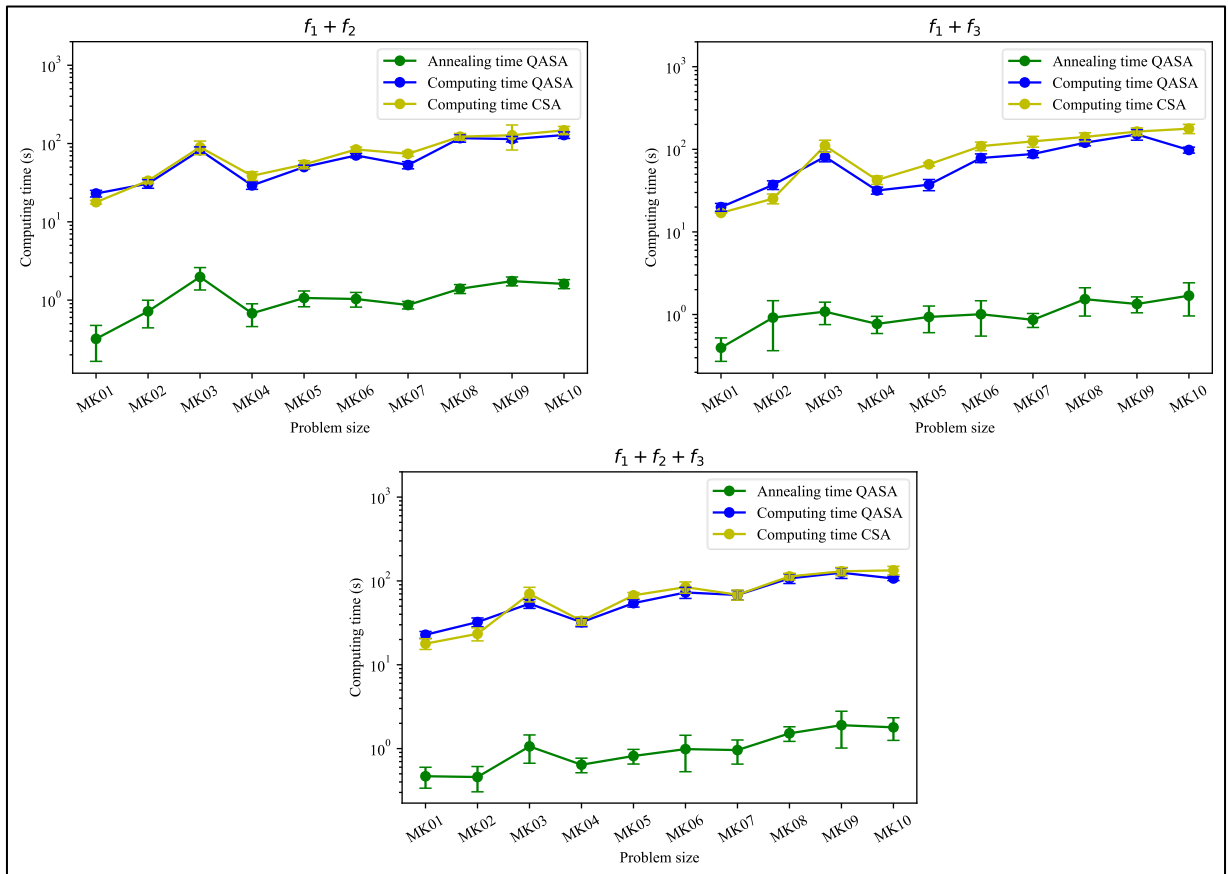

Figure 6: Comparison of computing times

In conclusion, the comparative analysis clearly demonstrates that QASA outperforms CSA across various metrics such as the C-metric, HRV, and computational efficiency. QASA proves to be highly effective in finding potential Pareto solutions, allowing for the identification of suitable solutions based on the weighting of objectives. While CSA may have some advantages in terms of computing time for smaller problem sizes, QASA excels as problem sizes increase. As a result, QASA shows great potential for industrial applications, particularly in handling larger planning problems within short timeframes while considering multiple objectives. To fully explore this industrial potential, further investigations are required. On the one hand it needs to be investigated whether the results of the study can also be extracted to large real-world application scenarios and on the other hand needs to be established the relationship between the selection of inputs (weights and solver parameters) and the dependency of Pareto points since foundation of the investigation were full factorial trials. This will enable the customization of parameters to directly align with the preferences and priorities of specific industrial operations. Additionally, there is a need to explore how the proposed approach can effectively map continuous time intervals to discrete time intervals. Overall, with its superior performance and the potential for customization, QASA represents a promising approach for industrial use, paving the way for efficient and effective decision-making in complex planning scenarios. However, it is important to note that practical statements about the suitability of the results cannot be made at this time. One of the reasons for this limitation is the formulation of the model. In practical applications, continuous times, such as those used for due dates, are prevalent, whereas the selected use case exclusively considers discrete time steps. To enable practical applications and provide meaningful insights, further investigations must be conducted, and methods for discretization need to be integrated into the calculation model. Additionally, validation using real planning data from industrial partners is essential before any conclusive assessments can be made. However, owing to its demonstrated superior performance, there is reason to anticipate that QASA possesses the potential to deliver substantial value, thereby equipping organizations with the capability to render efficient and effective decisions when confronted with intricate planning challenges.

## 5. Conclusion and Outlook

The outcome of process scheduling as part of the PPC has a direct impact on the operating costs of a manufacturing system. To minimize these costs, it is crucial to employ efficient planning algorithms that can solve complex problems while computing them within the shortest time intervals. However, due to the inherent complexity of such planning problems, heuristic planning approaches are commonly used. Nevertheless, these approaches have limitations in handling highly complex problems. As problem complexity increases, a trade-off between solution quality and computation time arises. Recent studies have introduced QA as a potential solution for solving complex assignment problems. However, these studies have been limited to small problem sizes and mono-criteria objectives. Consequently, they fail to meet the requirements of industrial applications, which demand consideration of multiple optimization objectives simultaneously. To address this research gap, this paper presents a QASA for the FJSSP. The mathematical formulation and framework for QASA are introduced, followed by a comparison with a CSA. The comparisons involve computing different problem instances considering various objective functions. Evaluation metrics are used to assess the Pareto fronts and computation time required for each algorithm. The results of the comparisons demonstrate that QASA outperforms CSA. QASA achieves qualitatively better results for each problem size, according to the C-metric in 100% of cases and according to HRV in 84% of cases. Additionally, QASA exhibits lower computation times for larger problem instances. Consequently, in the considered scenario, QA proves to be a suitable tool for computing FJSSP with multiple objectives. Furthermore, the diverse solution combinations on the Pareto front allow for the generation of different solutions based on preference, adaptable to the problem at hand.

These research findings provide a solid foundation for potential industrial applications. However, it is important to emphasize that establishing the relationships between parameter selection and planning results is crucial for the industrial use of this approach. For instance, in future work, deep learning methods will be employed to predict optimized parameter combinations based on historical planning data. Moreover, additional methods for variable reduction will be explored to handle the large number of variables for direct industrial use. Among other things, to adapt the approach to industrial scales. In addition, dynamic events will be included, as well as additional objectives such as minimizing tardiness. This comprehensive approach seeks to faithfully emulate the practical complexities and boundary conditions encountered by industrial companies. Thus, this study lays the groundwork for the evaluation of an industrial application of QASA in the future.

## Acknowledgment

This research was funded by the Ministerium für Wirtschaft, Verkehr, Landwirtschaft und Weinbau Rheinland-Pfalz - 4161-0023#2021/0002-0801 8401.0012.